\documentclass[aps,prb,twocolumn,a4paper,footinbib,longbibliography,showpacs,groupedaddress,floatfix]{revtex4-1}

\usepackage[dvipsnames]{xcolor}
\definecolor{linkcolor}{rgb}{0.3,0.3,1.0} 
\usepackage[pdftex,colorlinks=true, linkcolor= linkcolor, citecolor= linkcolor, urlcolor= linkcolor, hyperindex=true,hyperfigures=true]{hyperref} 

\usepackage{amssymb,amsmath,amsfonts}
\usepackage{graphicx}

\renewcommand{\vec}[1]{\mathbf{#1}}

\begin{document}

\title{Machine learning understands knotted polymers} 

\author{Anna Braghetto}
\affiliation{Department of Physics and Astronomy, University of Padova, 
Via Marzolo 8, I-35131 Padova, Italy}
\affiliation{INFN, Sezione di Padova, Via Marzolo 8, I-35131 Padova, Italy}

\author{Sumanta Kundu}
\affiliation{Department of Physics and Astronomy, University of Padova, 
Via Marzolo 8, I-35131 Padova, Italy}
\affiliation{INFN, Sezione di Padova, Via Marzolo 8, I-35131 Padova, Italy}

\author{Marco Baiesi}
\affiliation{Department of Physics and Astronomy, University of Padova, 
Via Marzolo 8, I-35131 Padova, Italy}
\affiliation{INFN, Sezione di Padova, Via Marzolo 8, I-35131 Padova, Italy}

\author{Enzo Orlandini}
\email{enzo.orlandini@unipd.it}
\affiliation{Department of Physics and Astronomy, University of Padova, 
Via Marzolo 8, I-35131 Padova, Italy}
\affiliation{INFN, Sezione di Padova, Via Marzolo 8, I-35131 Padova, Italy}

\begin{abstract}
Simulated configurations of flexible knotted rings confined inside a spherical cavity are fed into long-short term memory neural networks (LSTM NNs) designed to distinguish knot types. The results show that they perform well in knot recognition even if tested against flexible, strongly confined and therefore highly geometrically entangled rings. In agreement with the expectation that knots are delocalized in dense polymers, a suitable coarse-graining procedure on configurations boosts the performance of the LSTMs when knot identification is applied to rings much longer than those used for training. Notably, when the NNs fail, usually the wrong prediction still belongs to the same topological family of the correct one. The fact that the LSTMs are able to grasp some basic properties of the ring's topology is corroborated by a test on knot types not used for training. We also show that the choice of the NN architecture is important: simpler convolutional NNs do not perform so well. Finally, all results depend on the features used for input: surprisingly, coordinates or bond directions of the configurations provide the best accuracy to the NNs, even if they are not invariant under rotations (while the knot type is invariant). Other rotational invariant features we tested are based on distances, angles, and dihedral angles.
\end{abstract}

\maketitle

\section{Introduction}

Knotting is a non-local property of circular curves that
occurs naturally in sufficiently long swollen ring polymers~\cite{Sumners:Whittington:JPA:1988,Koniaris:prl:1991,orlandini:RMP:2007}
or globular ones~\cite{Janse-van-Rensburg&Whittington:1990:J-Phys-A,Mansfield:1994:Macromol,Virnau:JACS:2005,Baiesi:2011:PRL}.
Relevant biological examples are knots formed in long DNA strands
upon circularisation or in viral DNAs once densely packed in their capsids~\cite{Rybenkov:1993:PNAS,Shaw:1993:Science:8475384,ArsuagaPNAS2002,ArsuagaPNAS2005,MarenduzzoPNAS2009,Marenduzzo:PNAS:2013}.
In these cases knots can hinder elementary biological processes such as replication or transcription and evolution has provided specific enzymes, such as topoisomerases and recombinases, that facilitate the
cellular metabolisms by regulating the topological state of the DNA.

More generally, the occurrence of knots is known to affect the static, dynamical and rheological properties of the hosting polymer~\cite{Micheletti_PhRep2011,meluzzi2010biophysics,Mai:CurrOpColloid:2016,zhang2019effects} while, on the other hand, knotting and knot complexity strongly depend on both the physical properties of the polymer (for example, the stiffness~\cite{matthews2012effect,Poier:Macromol:2014,orlandini2016local,Coronel:SoftMatt:2017:bending}) and on whether the macromolecule is subject to external perturbations and constraints such as extensional forces and fields~\cite{Farago:2002:EPL,Bao:PRL:2003,Huang:JPCA:2007,Matthews:EPL:2010,Poier:Macromol:2014,Caraglio:PRL:2015,Najafi:2016,Narsimhan:MacroLett:2017:steady,Klotz:Macromol:2017:dynamics,Renner:SoftMatt:2015}, spatial confinement~\cite{Micheletti_PhRep2011,Tubiana:PRL:2011,Tang27092011,Rosa:PRL:2012,Micheletti:Macromol:2012,Micheletti:SoftMatt:2012,Plesa_et_al_NatNato_2016,SumaE2991,Liang_et_al_2012_Macro_Lett} and crowded environments~\cite{dAdamo:Macromol:2015}.

A main issue to face in studying knotting and knot complexity in polymeric systems is the
identification and classification of the knotted states in a very large set of
ring configurations (embeddings).
This has been mostly achieved by first projecting a given
configuration on a plane, then identifying and coding the resulting self-crossings and finally
feeding this information to build a topological invariant such as the Alexander
polynomial~\cite{Adams:1994,orlandini:RMP:2007}:
if the invariant differs for two different configurations, then they have different knotted states
(or do belong to two different knot types).
With this method, one can distinguish prime knots up to the minimal crossing number 7.
Since this approach originates from rigorous results in knot theory, it is exact.
However, the computation time of the whole procedure is known to increase with
the number of crossings $n_c$ formed after the projection and, if more sophisticated topological invariants
as the Jones and the HOMFLY polynomials~\cite{Adams:1994} are considered, the growth of time with $n_c$ is dramatic. 
This problem is particularly severe when either the length of the polymer $N$ is very large ~\cite{baiesi2012universal} or when the 3D configurations are geometrically badly organized in space, as for compact~\cite{Baiesi:2011:PRL} or isotropically confined rings~\cite{Micheletti:2006:J-Chem-Phys:16483240,Marenduzzo:PNAS:2013,Micheletti_PhRep2011}.

A way to curb this problem consists in performing local deformations that reduce as much as
possible the geometrical entanglement of the curve in 3D space
(and hence the number of crossings after projection) without changing its topological state~\cite{Koniaris&Muthukumar:1991a,Micheletti:2006:J-Chem-Phys:16483240, baiesi2014knotted}.
However, also these procedures are time consuming and there is no guarantee that they
can always decrease the number of crossings to the point at which the knot identification
based on an algebraic invariant is actually doable.

A different perspective on the problem of knot identification in circular polymers originates from the typical
goal of machine learning (ML) methods, namely to discover hidden patterns within a large set of complex data.
In the last years, the interest in these recognition techniques has grown dramatically also in physics~\cite{carleo2019machine}
where several machine learning algorithms have been introduced, tested, and applied to identify, for instance
new phase of matter, in particular topological phases~\cite{zhang2017quantum}, phase transitions~\cite{carrasquilla2017machine,venderley2018machine} and hidden symmetries and conservation laws~\cite{liu2022machine,liu2021machine}.

Very recently the use of ML supervised approaches based on neural networks (NNs) has been
recognized as a valuable tool in providing new insights into the mathematical problem of knot identification~\cite{vandans2020,davies2021advancing,kauffman2020rectangular,jejjala2019deep,gukov2021learning}. In Ref.~\cite{vandans2020} it was shown that a specific NN architecture, the long-short term memory (LSTM) NN, can learn
some global properties of the system that help identifying the knotted state, once trained on
sequences of bond directions along a semiflexible polymer ring of length N=100.

The work by Vandans et al~\cite{vandans2020} is the first attempt to use ML techniques to
identify and classify knotted states in polymer rings and its findings
have left open very interesting questions that we want to address here. These are the following:
i) Which configurational features or NN architectures are more effective in learning
global properties that best identify knotted states in polymer rings under different physical conditions?
ii) Unlike algebraic methods, the ML approach cannot identify with certainty a knotted state but
it can furnish a probability that a configuration has a given knot type. How does this degree of uncertainty
depend on the physical conditions of the rings and how topologically different is
the wrong prediction from the true one?
The last question is related to the more general problem of understanding
which properties of a knot type are learned and hence recognized by the NN.
Finally, the ML technique is probably not competitive in the task of recognizing knotted states
with respect to the algebraic invariant if the training procedure has to be carried out for any system under investigation.
One can, however, explore the possibility of exploiting the knowledge on knots acquired by the ML on a
given polymeric system (i.e. with a given contour length $N$ or density $\rho$) to
recognize knotted states in polymers with different properties. How effectively this can be done is another problem we plan to deal with here.

To accomplish the goals stated above we performed an extensive study of two NN architectures, the LSTM and the convolutional NN (CNN), see Section~\ref{s:NN} and Appendix.
These are applied to a model of flexible rings of different contour lengths $N$
and subject to different degrees of spherical confinement (see section~\ref{s:simul}).
For the training, several features have been tested and compared (see section ~\ref{s:input}). After the training of the LSTM, we look at its performance on unseen knotted configurations (section \ref {same_N}). This is done for a fixed polymer length $N$ and three different values of density. A comparison between the performance of ML trained with different features is performed too.
We next investigate how a LSTM, trained at a given length $N$ and density $\rho$ can be properly applied to recognize knotted states within rings of different length $N'$, after proper scaling of the features is performed (section \ref{diff_N}).

Since knot type recognition based on ML is not perfect, it is interesting to measure the degree of uncertainty of the NN by looking at how far, in the space of knot types, is the wrongly detected knotted state from the correct one. This is done in section~\ref{knot_families}. 
Section~\ref{conclusions} is devoted to discussion and conclusions.

\section{Simulations and Neural Networks}\label{simulations_and_NN}

\subsection{Simulations}
\label{s:simul}

To simulate fully flexible knotted ring chains, we use a bead-spring model where $N$ beads have coordinates $(\vec x_i)$ for $1 \le i \le N$. Each bead has mass $m$ and diameter $\sigma$. 
Excluded volume interaction between pairs $(i,j)$ of non-consecutive  beads along the chain is taken into account via the Weeks-Chandler Andersen (WCA) potential:
\begin{equation}
U_{\text{WCA}}(r) =
\left\{
\begin{aligned}
 & 4\epsilon\Big[\Big(\frac{\sigma}{r}\Big)^{12}-\Big(\frac{\sigma}{r}\Big)^{6}\Big]+\epsilon, & r \leq r_c \\
 & 0, & r > r_c
\end{aligned}
\right.
\end{equation}
where $\epsilon$ is the strength of the interaction, $r_c=2^{1/6}\sigma$ is the cutoff distance of the potential and $r = |\vec x_i-\vec x_j|$.
Consecutive beads along the chain are connected via spring with the finitely extensible nonlinear elastic (FENE) potential,
\begin{eqnarray}
U_{\text{FENE}}(r) &=& U_{\text{WCA}}(r)  \\
&&+\left\{
\begin{aligned}
 & -0.5kR_0^2\ln\Big[1-\Big(\frac{r}{R_0}\Big)^{2}\Big], & r \leq R_0 \\
 & \infty, & r > R_0
\end{aligned}
\right.\nonumber
\end{eqnarray}
where $r$ is the separation between bonded beads and the values $k=30\epsilon/\sigma^2$ and $R_0=1.6\sigma$ are chosen to avoid strand crossing~\cite{kremer_grest_1986,Kremer:1990:JCP}. 
This ensures that the knotted topology is preserved during the dynamical evolution of the system. With all these choices, each monomer in the system evolves through space in time following the Langevin dynamics at temperature $T$. The set of Langevin equations is solved numerically in NVE ensembles using a standard velocity-Verlet algorithm in LAMMPS~\cite{Plimpton:LAMMPS} with the damping factor $\gamma=1$ and a time step $dt=0.001\tau$, where $\tau=\sigma\sqrt{m/\epsilon}$ is the characteristic simulation time. Furthermore, we perform the simulations using $m=\sigma=\epsilon=T=1$.

Starting from an initial configuration of a polymer with a given knot type and chain length $N$, we generate compact configurations with  density $\rho=3N/(4\pi R^3)$ in a sphere of radius $R$ by using a spherical indenter that exerts a force on every bead which is directed towards the center of the sphere. The radius of the indenter, initially very large, is slowly decreased with time until the desired polymer density is reached. At this stage, we record the coordinates of the beads every $10^4$ simulation time steps (sampling period).

\begin{figure}[t!]
\centering
\includegraphics[width=0.98\linewidth]{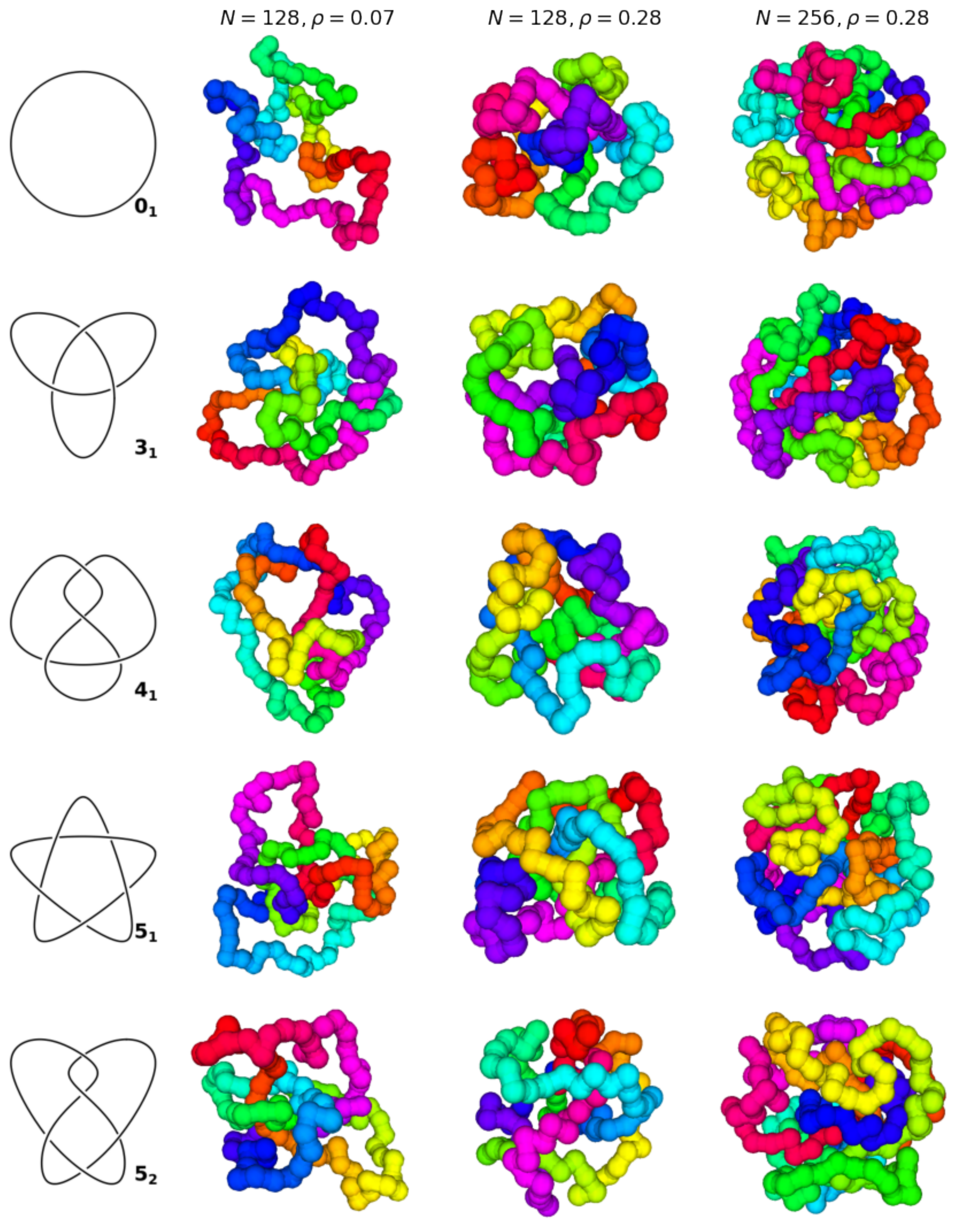}
\caption{Each row shows sampled configurations of knotted rings hosting one of the five different knots considered in this study, see the first column. The other columns refer to rings with: $N=128$ sampled at $\rho_1$, $N=128$ at $\rho_3$ and $N=256$ at $\rho_3$.
}
\label{fig:ex}
\end{figure}

In this paper, we initially consider five different types of knotted rings belonging to both torus and twist families: $0_1, 3_1, 5_1, 4_1,$ and $5_2$. For each knot type, we generated compact configurations for four different chain lengths $N$=64, 128, 256, and 512; and for each chain length for three different densities $\rho_1=0.07$, $\rho_2=0.14$, $\rho_3=0.28$. Some typical configurations are illustrated in Fig.\ \ref{fig:ex}.

\subsection{Input features}
\label{s:input}

\begin{figure}[t!]
\centering
\includegraphics[width=0.98\linewidth]{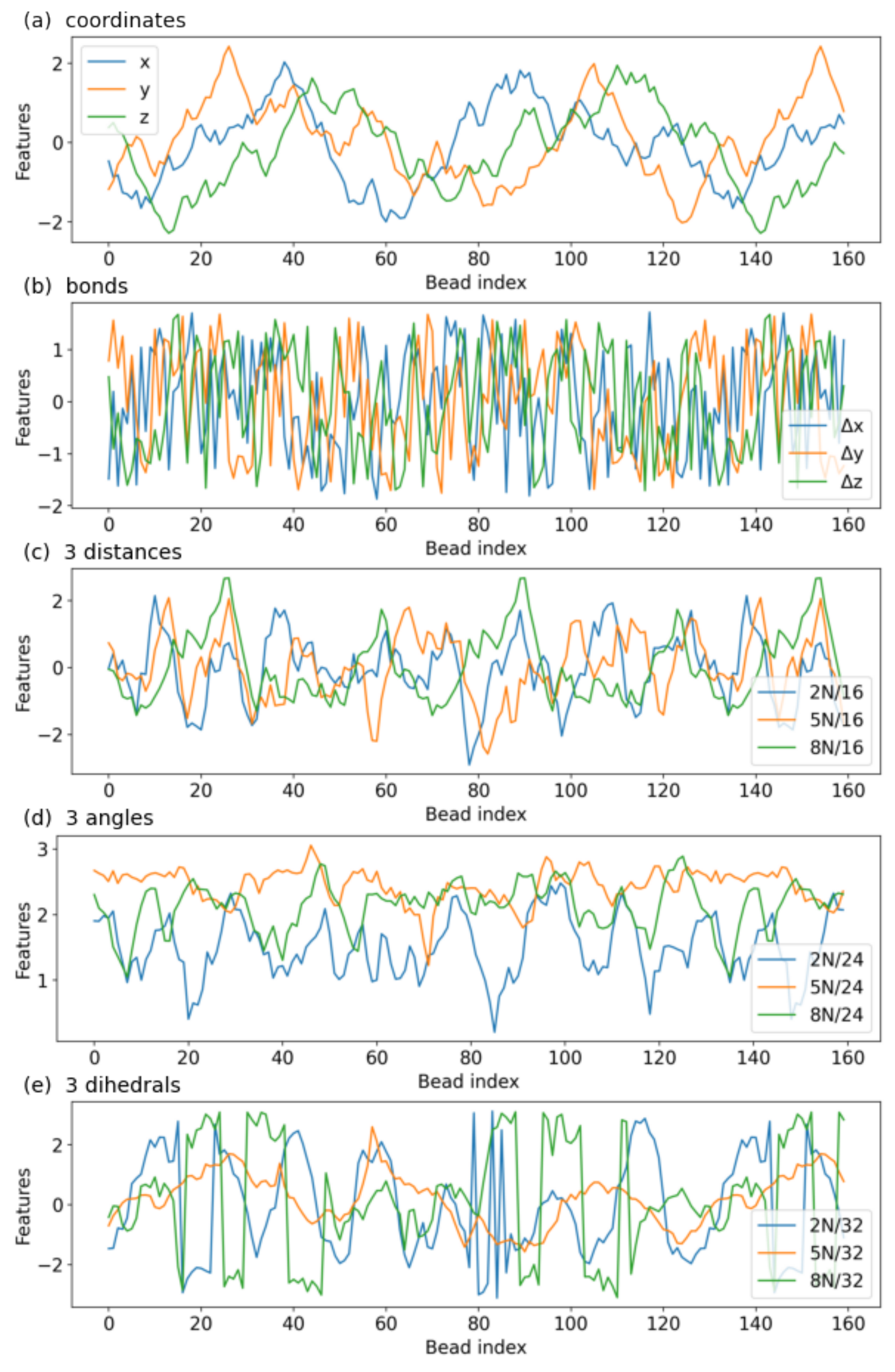}
\caption{Example of input features for a configuration with knot $3_1$, $N=128$, $\rho=\rho_3=0.28$: (a) coordinates, (b) bond directions, (c) 3 distances, (d) 3 angles, and (e) 3 dihedral angles. For the last three cases, the legend shows the values of the curvilinear distance $s$.
}
\label{fig:feat}
\end{figure}

For each sampled ring, centered in its center of mass, we computed different sets of features. Examples of input patterns from these features, described below, are shown in Fig.~\ref{fig:feat}. Because of the circular topology, it is understood that $N+1=1$ (periodic boundary conditions).

Perhaps the most natural features to consider are the coordinates $(\vec x_i)$ of the ring's beads and the bond directions $(\vec x_{i+1}-\vec x_i)$. In both cases, there are $3 \times N$ values to be stored in the ML. Note that these features are not invariant under rotations of the configuration, an operation that, on the other hand, does not change the knot type. 

To check whether this property can help improving the ML, we considered also sets of features that are rotational invariant. One of these is the set of $N$ distances between monomers separated along the ring by an arc length $s$,
\begin{equation}
 d_i(s) = \left|
 \vec x_{i+s} - \vec x_i
 \right|\,.
\end{equation}
By considering {$s=(2,5,8)\times N/16$} we collected $3\times N$ features for each ring that we stored in a matrix termed ``3 distances".
We choose $s\sim N$ for two main reasons: (i) knotting is a global property; (ii) in sufficiently dense conditions it is believed that the size of the knotted regions is proportional to $N$ (knot delocalization)~\cite{Marcone:PRE:2007,Tubiana:PRL:2011,Marenduzzo:PNAS:2013}. 

Similarly, we considered angles between three monomers, with pairs separated by arc length $s$,
\begin{equation}
 \theta_i(s) = \arccos\left[
 (\vec x_{i+s} - \vec x_i)
 \cdot
 (\vec x_{i+2s} - \vec x_{i+s})
 \right]
\end{equation}
and arrange them in a $3\times N$ matrix ``3 angles", by considering in this case {$s=(2,5,8)\times N/24$}.

Finally, we computed also the dihedral angle $\varphi_i(s)$ between four monomers, $\vec r_1 = \vec x_i$, $\vec r_2 = \vec x_{i+s}$, $\vec r_3 = \vec x_{i+2s}$ and $\vec r_4 = \vec x_{i+3s}$, namely the angle between the plane with points $(\vec r_1,\vec r_2,\vec r_3)$ and the plane with $(\vec r_2,\vec r_3,\vec r_4)$:
\begin{equation}
\begin{split}
 \varphi_i(s) &= \arctan \left[
 \frac{\vec u\cdot \vec w}{\vec v\cdot \vec w}
 \right]\\
 \vec v &= \frac{(\vec r_2 - \vec r_1)\times (\vec r_3-\vec r_2)}{|| (\vec r_2 - \vec r_1)\times (\vec r_3-\vec r_2) ||}\\
 \vec w &= \frac{(\vec r_3 - \vec r_2)\times (\vec r_4-\vec r_3)}{|| (\vec r_3 - \vec r_2)\times (\vec r_4-\vec r_3)||}\\
 \vec u &= \frac{(\vec r_3 - \vec r_2)\times \vec v}{||(\vec r_3 - \vec r_2)\times \vec v||}.
\end{split}
\end{equation}
The resulting $3\times N$ matrix, ``3 dihedrals", is then obtained by considering {$s=(2,5,8)\times N/32$}. These dihedral angles are indicators of the chirality of the chain path over long curvilinear scales $s\sim N$. Compared to the set of distances and angles, dihedrals are then expected to better describe the geometrical winding needed to generate a given knot type along a close chain.

Note that the scale-dependent features (i.e., coordinates, bonds, and distances) need to be normalized before the training: this is done by subtracting the average of their values along the ring and dividing the result by the corresponding standard deviation.

All features are periodic along the chain and the choice of the first monomer is arbitrary. Any other monomer would have generated a different cut in the sequences.
To mitigate the effect of this choice and feed NNs with a sequence that contains some information on this periodicity, all the $3\times N$ feature matrices were padded by attaching at position $N+1$ a copy $3 \times 32$ of the features for $1\le i \le N_P=32$. Hence, an actual input data sample is $3\times (N+N_P)$ matrix.

\begin{figure}[t!]
\centering
\includegraphics[width=0.98\linewidth]{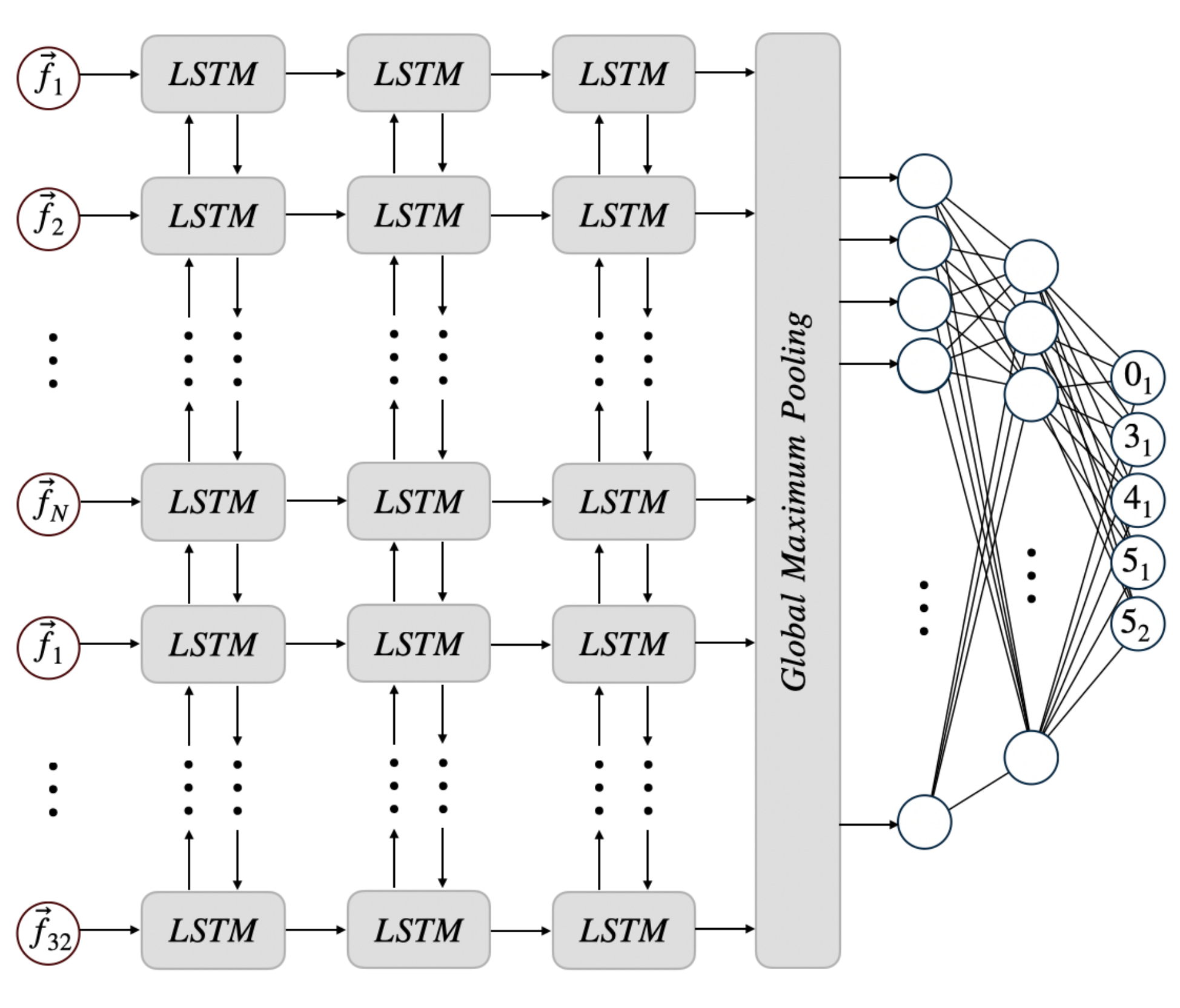}
\caption{Sketch of the unrolled bidirectional LSTM used in this work. The input sequence, containing $N+N_P$ 3-dimensional vectors, is processed in both directions (i.e., bidirectional) by the first LSTM layer, whose cell has 64 units along each direction, hence it has 128 units in total. The processed sequence of features is then fed to 2 subsequent LSTM layers, whose cells have 128 units each. We then perform global maximum pooling, obtaining a vector with $F=128$ neurons. The $F=128$ neurons start a feed-forward neural network including other two dense layers with 64 and 5 units each. In the last dense layer, we apply a softmax function, while we apply leaky ReLU activation functions to the other layers. In addition, as a common practice to reduce overfitting, in the feed-forward component we include a dropout procedure with a rate of $20\%$.
}
\label{fig:lstm}
\end{figure}

\subsection{Neural Networks}
\label{s:NN}

While in the Appendix we discuss also a ML algorithm for knot identification based on convolutional neural networks (CNNs), most of the findings reported here are based on a
bidirectional Long-Short Term Memory (LSTM)~\cite{lstm}, a NN that we built with 3 recurrent layers, including 128 units each, see Fig.~\ref{fig:lstm}.
CNN and LSTM are artificial neural networks that are often implemented to analyze either time series or any (multi-dimensional) signal measured at a given time. In these cases, the NNs are able to extract relevant patterns in the analyzed sequences and perform tasks such as classification, regression etc. In our approach, time series are replaced by polymer configurations and the signal is represented by a set of features whose dimension corresponds to the number of features in each set (i.e., 3) while the instants of the time series are replaced by the labels of the bead along the rings. Note that CNNs and LSTMs belong to different classes of neural networks. While CNNs are feed-forward neural networks originally developed to process images, LSTMs belong to the recurrent neural network class, built to deal with sequences such as time series or sentences. In particular, LSTMs are made up of a cell that learns long and short-range correlations in the sequences (e.g., among beads in the ring) thanks to its components, called gates (i.e., input, forget, and output), which drive the information flow. Under these settings, we expect that LSTM will be able to discriminate among rings with a given knot type better than CNN (see Appendix for a comparison between the two NNs). Moreover, we resort to the bidirectional version of the LSTM since rings (circular chains) do not have a preferred direction along the chain.

\begin{figure*}[tb!]
\centering
\includegraphics[width=.63\textwidth]{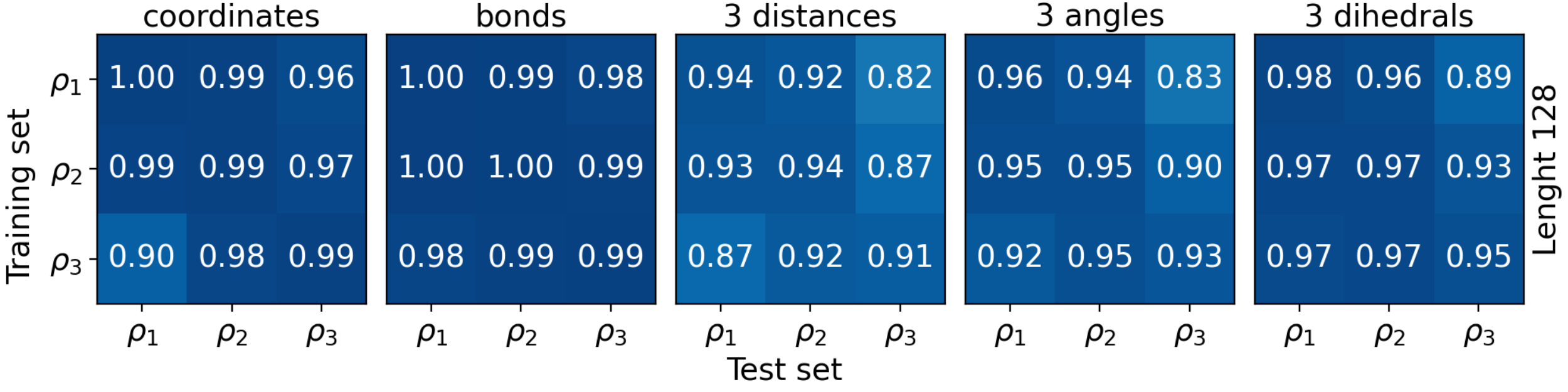}
\caption{Test accuracy on polymers with length $N'=128$, as those used for training. Each slot corresponds to a given density of the configurations in the training set (left axis) and the density of test configurations (bottom). The five squares correspond to different features used for training.
}
\label{fig:acc}
\end{figure*}

 The training procedure we considered is the following.
 The several layers of the LSTM NN are set up to generate an output of $F=128$ units, hence a matrix of $F\times  (N + N_P)$ values is produced. To remove the dependence on $N$, we performed a global maximum pooling procedure along the padded length $N+ N_P$ of the input chain (see Fig.~\ref{fig:lstm}). The values obtained from the global maximum pooling are assigned to the neurons in the first layer of the feed-forward component of the NN. This is connected to another dense layer with $64$ units, which is feeding the output layer. Since we are dealing with a multi-classification scheme with $5$ classes (i.e. $5$ knot types), the output layer has $5$ units, to which we apply the softmax function. 
Therefore, the neural network takes in input a feature matrix with shape $3\times  (N + N_P)$, where $N$ can assume any integer value, and returns a vector with 5 entries, each one representing the probability $p_k$ that the input sample belongs to the knot type  $k$.

With this procedure, we trained different models, one for each set of features, using different training sets (i.e. ring configurations), each one sampled at the three given densities $\rho_1,\rho_2$, and $\rho_3$. For each density, we collected $6\times 10^5$ samples per knot type that we split into a training ($80\%$) and a validation ($20\%$) set. To monitor the possible overfitting on the data, the splitting is carried out by taking into account the temporal order of the simulations. 
More precisely, for each knot type, three independent runs generate $2\times 10^5$ configurations each, and the training set collects the first $1.6\times 10^5$ ones from each run. The validation set includes the remaining configurations. 

Although the architecture of the LSTM allows in principle a training with configurations sampled at different lengths $N$, we limit our training set to rings with $N=128$. 
To increase the performance of the NN and to avoid overfitting during learning, we also monitored the evolution of training by looking at the learning rate decay and using the early stopping procedure. The learning rate decay allows reducing the learning rate parameters of a given factor ($1/2$ in this work) when we do not observe any improvement of the validation loss after a given number of epochs (10 in this work). The early stopping procedure allows to avoid overfitting by stopping the training procedure when the validation loss starts to increase for a given number of epochs (30 in this work). If a learning procedure is interrupted in this way, the NN weights are recovered from 30 epochs earlier than stopping.

\begin{figure*}[tb!]
\centering
\includegraphics[width=.63\textwidth]{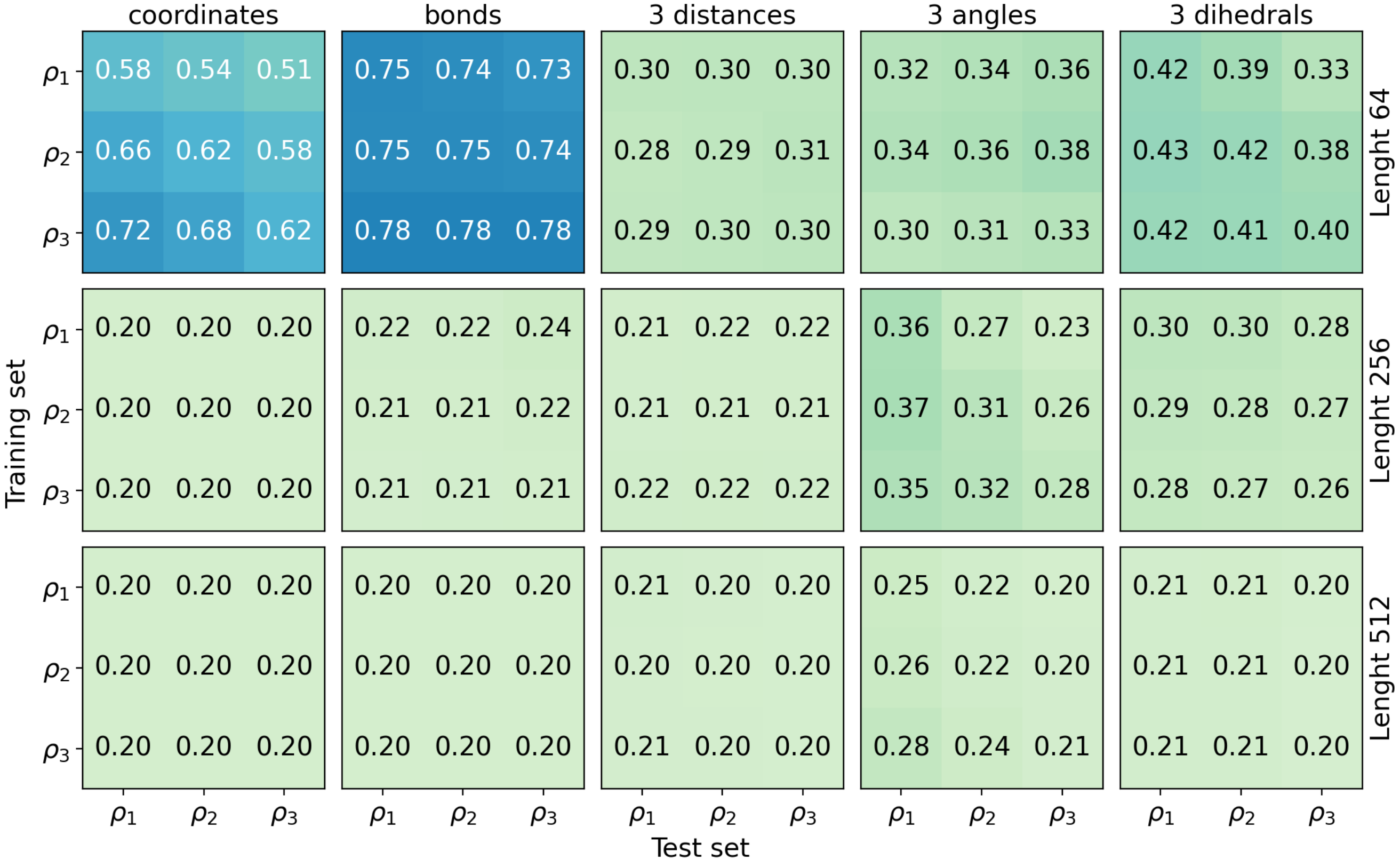}
\vspace{-1mm}
\caption{Test accuracy as in Fig.~\ref{fig:acc} but for configurations with $N'=64$ (first row), $256$ (middle row), and $512$ (bottom row).
}
\label{fig:acc2}
\end{figure*}

\section{Knot classification by LSTM}\label{Knot_classification}
After training the LSTM on rings with $N=128$ and using the different sets of features described in section~\ref{s:input} (training sets), we tested its ability to recognize knotted states among the 5 possible ones (i.e. $0_1,3_1,4_1,5_1$ and $5_2$) towards newly generated configurations  (testing set).
For this purpose, from an independent run we simulated additional $5\times 10^5$ configurations per knot type and given density $\rho\in \{\rho_1,\rho_2,\rho_3\}$ and for rings of contour length $N'=64,128,256$ and $512$.
As an index of the model quality, we used the accuracy $A$, namely the fraction of correctly classified samples. 

\subsection{Recognizing knots in confined rings at fixed $N$}
\label{same_N}

First, we tested the LSTM on rings having the same contour length as the training ones ($N'=128$).
Considering the three simulated densities, this amounts to nine test sets per feature. The results are summarized as a $3\times 3$ matrix representing the accuracy obtained by the LSTM trained with the feature indicated on its top (see Fig.~\ref{fig:acc}). The training density is indicated on the left axis while the test density is on the bottom one.

One can notice that the LSTM trained on features such as coordinates, bonds, and dihedral angles display the best accuracy (value of A close to $1$), though in all tables we note a good average accuracy $A$. As expected, the largest values of $A$ are along the diagonal, where the test density matches the training density. It is worth noticing that the off-diagonal $A$ values are also quite large: $0.98 \le A\le 1$ for the bond features (best case) and $0.82 \le A\le 0.93$ for 3 distances (worst case). These results show that the LSTM NN can recognize, with a very good degree of accuracy, knotted states in rings sampled at densities also different from the one of the configurations used for training. 
Quite remarkably, the accuracy is sufficiently high even for the most extreme case where knot recognition is carried out on testing rings that are much denser ($\rho=\rho_3 = 0.28$) and hence more crumpled than the training ones ($\rho_1 = 0.07$), see top right squares in each table. 

This also suggests that the patterns learned by the ML are sufficiently robust with respect to the large amount of geometrical entanglement that is 
present in highly confined rings.

\subsection{Recognizing knots in rings with $N'\ne N$}
\label{diff_N}
We now ask to which extent the NN, trained on rings of a given length (here $N=128$), can recognize knotted patterns in shorter ($N'=64$) or longer ($N'=256,512$) rings. 
For $N'=64$ the corresponding accuracy $A$ is reported, for the five different features considered, in the $3\times 3$ tables of the top row of Fig.~\ref{fig:acc2}: one can readily see that, apart from a partial achievement observed for the coordinates ($0.51<A <0.72$) and bonds ($0.73<A <0.78$) features, the accuracy is in general unsatisfactory. Note that in Vandans et al~\cite{vandans2020} the tests performed on shorter rings seem more satisfactory than those reported here. However, in their study, the knotted rings were semiflexible and hence less geometrically  entangled than those considered here. Moreover, their ML was trained on rings of length $N=100$ and the tests were performed on rings of length $N'=80,60$. In any case, from both studies, it emerges the picture that $A$ tends to worsen when $N'$ becomes increasingly smaller than $N$.

The degree of inaccuracy in knot recognition is even more dramatic if the testing rings are much longer, namely, $N'=256$ and $512$, see the second and third row of Fig.~\ref{fig:acc2}. In these cases the prediction of the NN is comparable to a random choice of the knotted state between the five possible ones ($A \sim 0.2$).

A way to make knot recognition less sensitive to the ring's length is based on the assumption that knotting, being a global property of a closed curve, should be mildly affected by a sequence of local deformations along the chain that scale the original length of the configuration $N'$ to that used for training.

\begin{figure*}[tbh!]
\vspace{3mm}
\centering
\includegraphics[width=.63\textwidth]{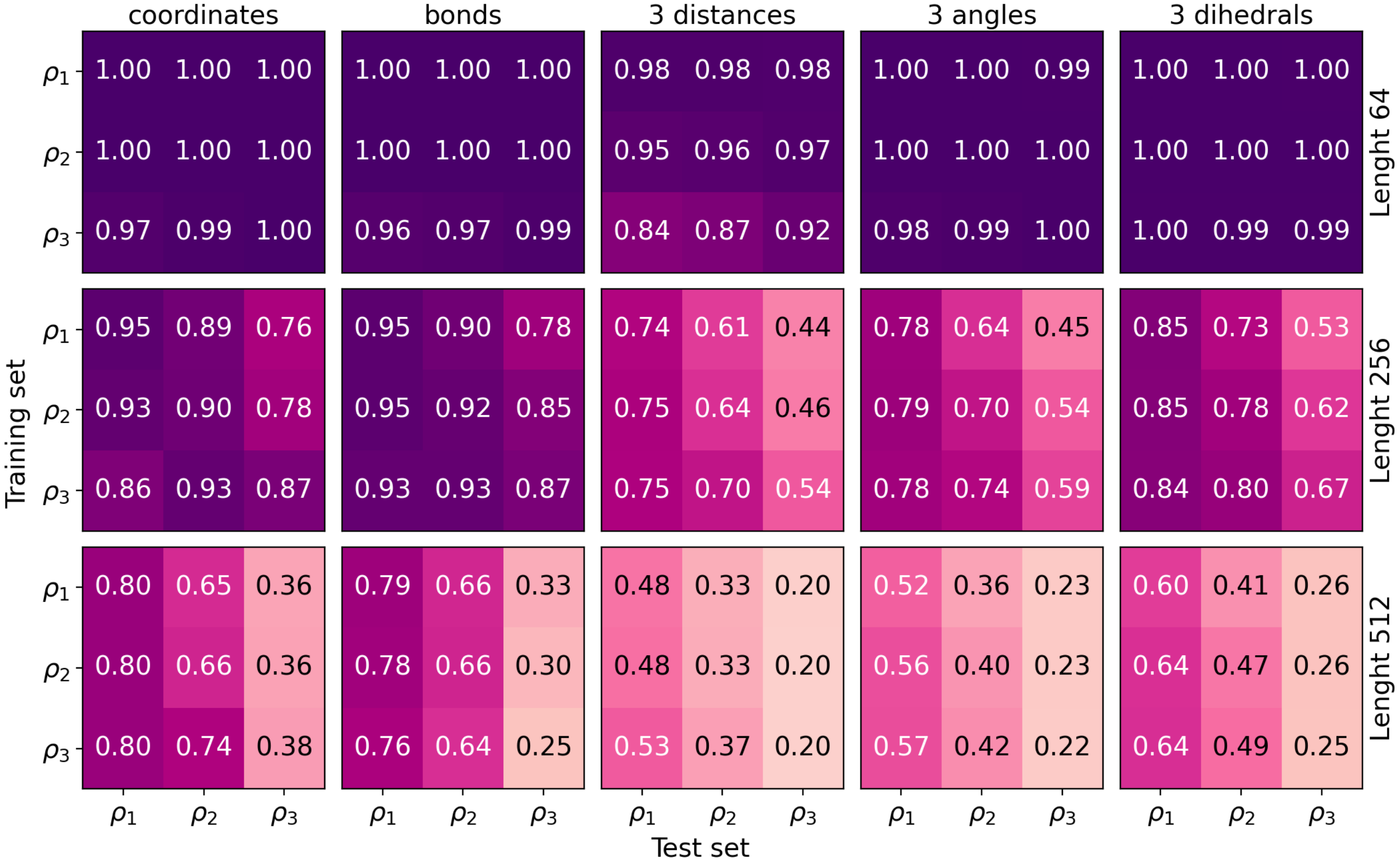}
\vspace{-1mm}
\caption{Test accuracy as in Fig.~\ref{fig:acc2}, but after the rescaling of the chain length $N'$ to $N=128$ as described in the text.
}
\label{fig:acc3}
\end{figure*}

According to this idea, we performed a prior transformation on the testing rings as follows:
For shorter rings ($N'=64$) we considered a fine graining procedure where an additional bead is inserted between each pair of consecutive beads of the original configuration, namely
\begin{align}
 \vec x_i &\to \vec X_{2i-1}\nonumber\\
 \frac 1 2(\vec x_i+\vec x_{i+1}) &\to \vec X_{2i}
\end{align}
with $1\le i \le N'$.
Rings with $N'=256$ are instead coarse-grained into a ring of length $N=128$ by replacing each pair of contiguous beads (bond) with its midpoint:
\begin{align}
 \frac 1 2(\vec x_{2i-1}+\vec x_{2i}) &\to \vec X_{i}
\end{align}
For the case with $N'=512 = 4\times N$ the coarse-graining is iterated twice. 
In this way all the testing rings of length $N'$ are converted into a set of coordinates $\vec X_i$ with $1\le i \le N=128$ and all the features are computed with respect to these new coordinates.

The accuracy of the NNs tested on the rescaled configurations are reported, for different features, in Fig.~\ref{fig:acc3}: the improvement after the rescaling is relevant. This is especially true when the test is performed on shorter rings ($N'=64$) (first row) with the best accuracy that is found when the features considered are the 3 dihedral angles ($0.99 \le A \le 1$). It is worth noticing that the computed efficiency is essentially independent of the density at which rings have been sampled. The improvement upon rescaling is substantial also for longer rings. For instance, for $N'=256$ and using coordinates as features, the accuracy increases from $0.20$ (random) to values larger than $0.75$ and up to $0.95$ when training and test sets have both been sampled at $\rho_1=0.07$. The improvement is still quite significant also for the case $N'=512$, i.e., when the testing rings are four times longer than the training ones. For instance, the value $A=0.80$ of the left column of the coordinates matrix means that, by performing a training with knotted rings of length $N=128$ and $\rho_1 = 0.07$, the ML is able to recognize with $80\%$ of accuracy the correct knotted state among the 5 possible ones in rings of length $N'=512$, irrespective of the density at which the testing rings have been sampled.

The fact that, for all the features considered, the accuracy is higher when training is performed at the smallest density $\rho_1$, suggests that in long knotted rings ($N'=512$) the recognition of the knotted pattern is more effective when rescaling is performed on mildly confined rings.

For denser polymers, the improvement is still noticeable, though. This can be qualitatively understood by noticing that physically knotted regions in compact or densely packed rings are known to have an average length that grows with $N$ ~\cite{Tubiana:PRL:2011,Tubiana:PTPS:2011,Marcone:PRE:2007,Marenduzzo:PNAS:2013}. One can then argue that coarse-graining rings from $N'$ to $N$ effectively rescales also the size $L'$ of the original knotted region to the one, $L$, expected in rings of length $N$. This argument is rather qualitative since it relies on the scaling behavior $L \sim N$ that is valid in the large $N$ limit and hence does not take into account finite-size effects. Another possible drawback of the coarse-graining procedure is that, by replacing pairs of coordinates with their midpoints, some information can be lost, or, more importantly, the topology of the original ring can change by an effective strand passage. Given all these potential problems, the good $A$'s obtained for rings of length $N'=512$ for trained ML at $N=128$ is rather intriguing.

\begin{figure*}[tbh!]
\centering
\includegraphics[width=.9\textwidth]{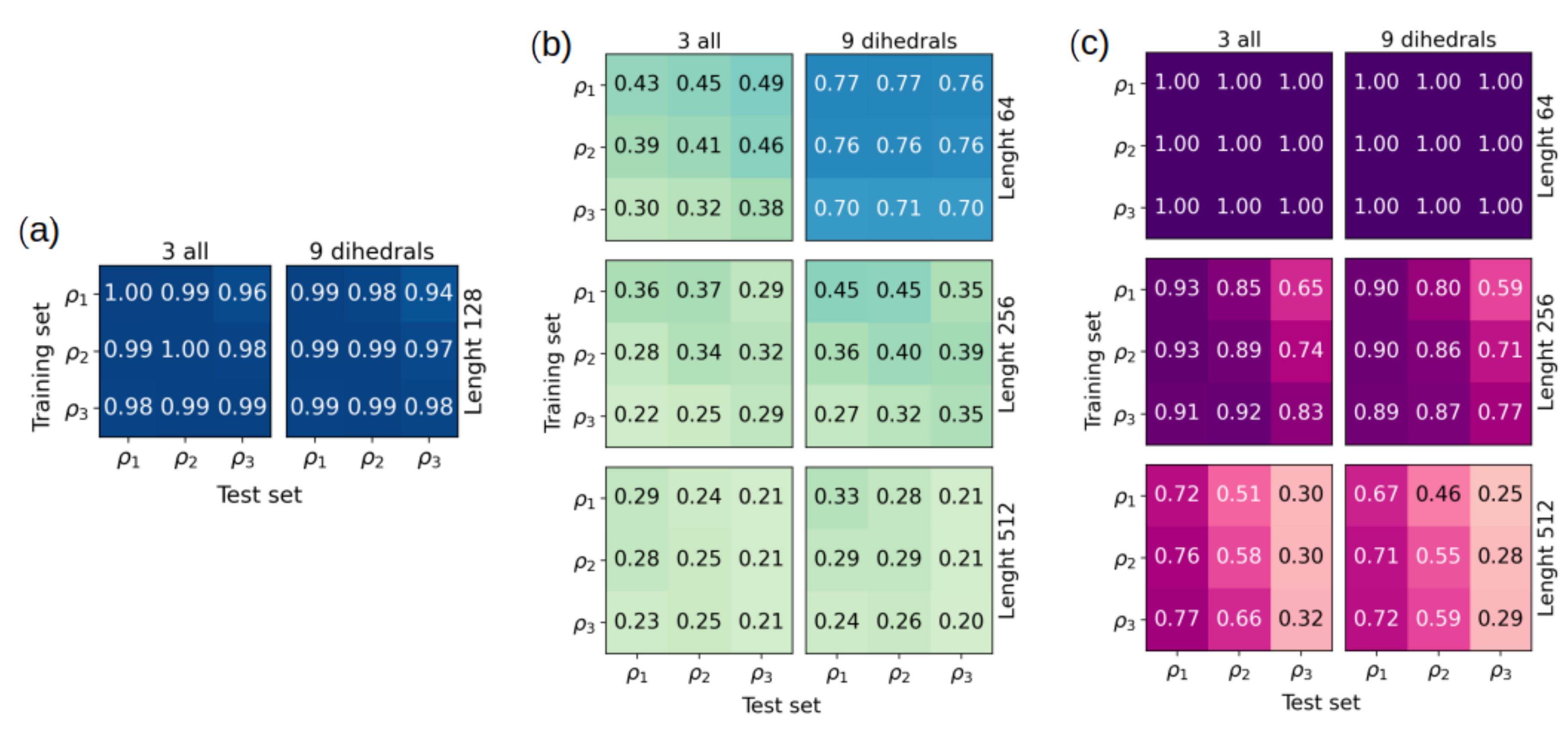}
\caption{LSTM accuracy after its training with $9\times  (N + N_P)$ rotational invariant features for rings of length $N=128$, sampled at different densities. ``3 all" stands for a combination of 3 distances, 3 angles, and 3 dihedrals; ``9 dihedrals" uses 9 generalized dihedral angles.
(a) Test accuracy for $N'=128$ for different training densities and at different test densities. Training at the smaller density $\rho_1$ and testing at the highest density $\rho_3$ appears to be the worse case ($A=0.94$ for ``3 all" and $A=0.96$ for ``9 dihedrals"), however, in all cases the accuracy is high.
(b) The same for $N'=64, 256$, and $512$. Generalization to different lengths does not work in this case but can be fixed: (c) shows the improved accuracy for NN tested with rescaled configurations.
}
\label{fig:acc9}
\vspace{-1mm}
\end{figure*}

\subsection{Increasing the space of rotational invariant features }

A quite intriguing result is that LSTMs trained with features 
such as coordinates and bonds often perform better than those based on rotational invariant features such as distances, angles, and dihedrals. A possible explanation is that the choice of limiting these features to a set of $3\times N$ values is not optimal to achieve a faithful description of a 3D configuration. Possibly, some information is lost during the conversion from coordinates to these processed features.

We then tested the accuracy of LSTMs when larger input matrices of rotational invariant features are considered. For instance, one can build a set of $9\times N$ matrices of mixed features (``3-all") by simply appending the $3\times N$ values of the distances, angles, and dihedrals features each computed with the three step size
$s=\frac{N}{32}\times(2,5,8)$.
We also built a ``9 dihedrals" set with $9$ different values of the generalized dihedral angles, computed along the rings using steps of size
$ s=\frac{N}{32}\times \left(\frac{32}{N}, 1,2,3,4,5,6,7,8\right )$.
Note that with this choice of step size the $3\times N$ values previously used for the ``3-dihedral" set are included in this larger set. 

The corresponding results are summarized in Fig.~\ref{fig:acc9}.
One can notice that when $N'=N$ (Fig.~\ref{fig:acc9}a), the accuracy is greatly improved with respect to those obtained with the $3\times N$ rotational invariant sets (see last three matrices of Fig.~\ref{fig:acc}). More importantly, the values of $A$ are now comparable or larger than those observed when coordinates or bonds are considered (see the two leftmost matrices of Fig.~\ref{fig:acc}).

For $N' \ne N$ and with no prior rescaling, the accuracy is still poor but far better than those obtained by relying only on $3\times N$ rotational invariant features (compare Fig.~\ref{fig:acc9}(b) with Fig.~\ref{fig:acc2}). 

With rescaling, there is a $100\%$ of accuracy for shorter rings ($N'=64$) and a great improvement, both for $N'=128$ and $512$, with respect to all the $3\times N$ sets of rotational invariant features (see Fig.~\ref{fig:acc9}(c) and Fig.~\ref{fig:acc3}). In particular, the new accuracies are now comparable with those based on either coordinates or bonds.

\begin{figure*}[t!]
\centering
\includegraphics[width=0.96\textwidth]{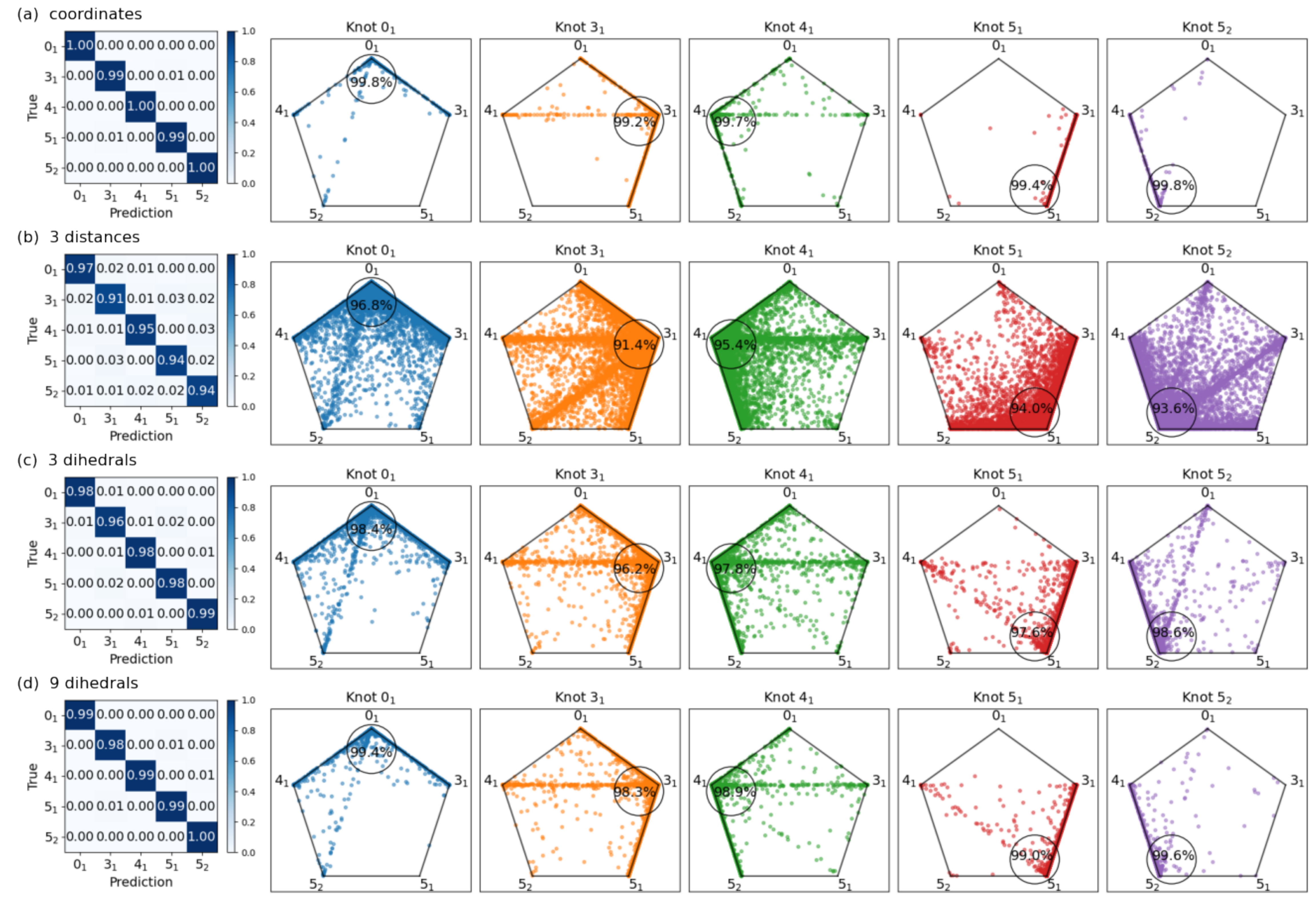}
\vskip 0mm
\caption{Each row corresponds to a feature and contains the confusion matrix and the knot space for each of the five considered knots, for simulations with $N=N'=128$ and $\rho_1=0.07$ for train and test. We show results for (a) coordinates (they are similar for bond directions), 3 distances (they are similar for 3 angles), 3 dihedral angles, and 9 dihedral angles. } 
\label{fig:penta007}
\end{figure*}

\section{Machine learning families of knots}\label{knot_families}

Since knot recognition based on ML carries a given degree of uncertainty, it is interesting to understand how badly the ML can be mistaken and how its degree of confusion depends either on the analyzed knotted state  or on the adopted training procedure (e.g. set of features employed). For instance, one can ask how far is, in the topological complexity ladder, the wrongly detected knotted state from the correct one. 

This issue can be quantitatively addressed by looking at the
output of our NNs, namely a softmax function with $1 \le k \le 5$ values $p_k$ representing the predicted probability that the input corresponds to a knot of type $k$. For visualization, this five-dimensional vector can be embedded into a two-dimensional region by choosing five unit vectors $\vec v_k$ on the plane, each pointing to a different vertex of a regular pentagon. In this way the predictions $(p_k)$ are mapped to the region
\begin{equation}
 \vec z = \sum_k p_k \vec v_k 
\end{equation}
within the pentagon.
Note that this is a sort of generalized simplex in which an injection to a space of a smaller dimension is achieved.

This representation helps visualizing the paths of mistaken knots mostly followed by the ML when some uncertainty in the prediction $(p_k)$ is present. 
Here we focus on the simplest case of $N'=N=128$ and when the training and testing sets have been both sampled at $\rho=\rho_1=0.07$.

In Fig.~\ref{fig:penta007} one can notice lines joining the corner of the pentagon corresponding to the correct knot $k_c$ with another corner representing one of the 4 possible wrong knot types. These lines are the simplest path of mistakes where there is a non-zero uncertainty on the correct knot type $k_c$ (i.e. $p_{k_c}\ne 1$) and there emerges a second knot type $k$ having $p_k\ne 0$ while the remaining $p_{k'}\approx 0$ (i.e. $p_{k_c}+p_k \simeq 1$). Clearly, points in the interior of the pentagon correspond to cases in which the degree of uncertainty is shared by more than one wrong knot type. A more quantitative information is provided through the confusion matrices of the results, see the first column of Fig.~\ref{fig:penta007}.

Let us first focus on the coordinates, from which we expect smaller uncertainties in the performance. Indeed, the corresponding 5 pentagons (one for each knot type $k_c$) mostly display a clean aggregation of points along the straight segments connecting the different knot types (see Fig.~\ref{fig:penta007}(a)). This means that, if the ML gets confused, it is mostly between $k_c$ and another knot $k$. More importantly, the path of uncertainties is along knots belonging to the same knot family. 
For instance, if $k_c$ is the torus knot $5_1$~\footnote{A torus knot is a knot that can be embedded on the surface of a torus} the ML recognizes $k_c$ in $99.4\%$ of the rings while for most of the remaining cases the choice is on $3_1$, the simplest knot of the torus family. Similarly, if $k_c$ is the twist knot $5_2$~\footnote{A twist knot can be created by twisting an unknotted ring and linking the end loops}, the prediction is correct for $99.8\%$ of the tested rings while most of the confusion is with $4_1$, the knot closest to $5_2$ within the family of twist knots. Note  that also the segment from $5_2$ to the unknot is populated indicating that ML can get confused within the twist knot with 5 crossings and the unknot. This can be understood by recalling that all twist knots have unknotting number one, namely, they can be converted to the unknot by a single strand passage~\cite{Adams:1994}.

Finally, if we look at $k_c = 3_1$ there is a $99.2\%$ of correct predictions while the source of confusion is due to the unknot, the $4_1$ (twist), and the $5_1$ (torus) knots. Apart from the obvious connection with the unknot the fact that ML mostly gets confused with either $4_1$ or $5_1$, in agreement with the topological property of $3_1$ being both a twist and a torus knot. 

Similar patterns are observed also for features other than the coordinates, although in cases such as the ``3-distances" (Fig.~\ref{fig:penta007}(b)) and the ``3-dihedrals" (Fig.~\ref{fig:penta007}(c)) there is a larger uncertainty  that is shown  by a non-negligible spread of the points within the internal regions of the pentagon that do not belong to  the segments joining the 5 corners. 
It is however important to stress that in all the scatter plots of Fig.~\ref{fig:penta007} the visible points represent the minority of cases, as most points are stacked on the top of the corner representing $k_c$. 

The situation improves if the set of 9 dihedrals is considered instead, see [Fig.~\ref{fig:penta007}(d)].

Similar results are found also at $\rho=\rho_3$, but with a wider spreading of points (not shown)
since higher densities, configurations are geometrically more entangled and knot recognition is more difficult to achieve.

In summary, the plots in the pentagon space are in all cases compatible with, and complement the information one can also get from the confusion matrices.
The results shown in Fig.~\ref{fig:penta007}
suggest that a NN, even when mistaken, is grasping some general topological properties of the observed configuration, such as its belonging to a given class or family of knots. 

\subsection{Dealing with ever seen knots}
\begin{figure}[tb!]
\centering
\includegraphics[width=.33\textwidth]{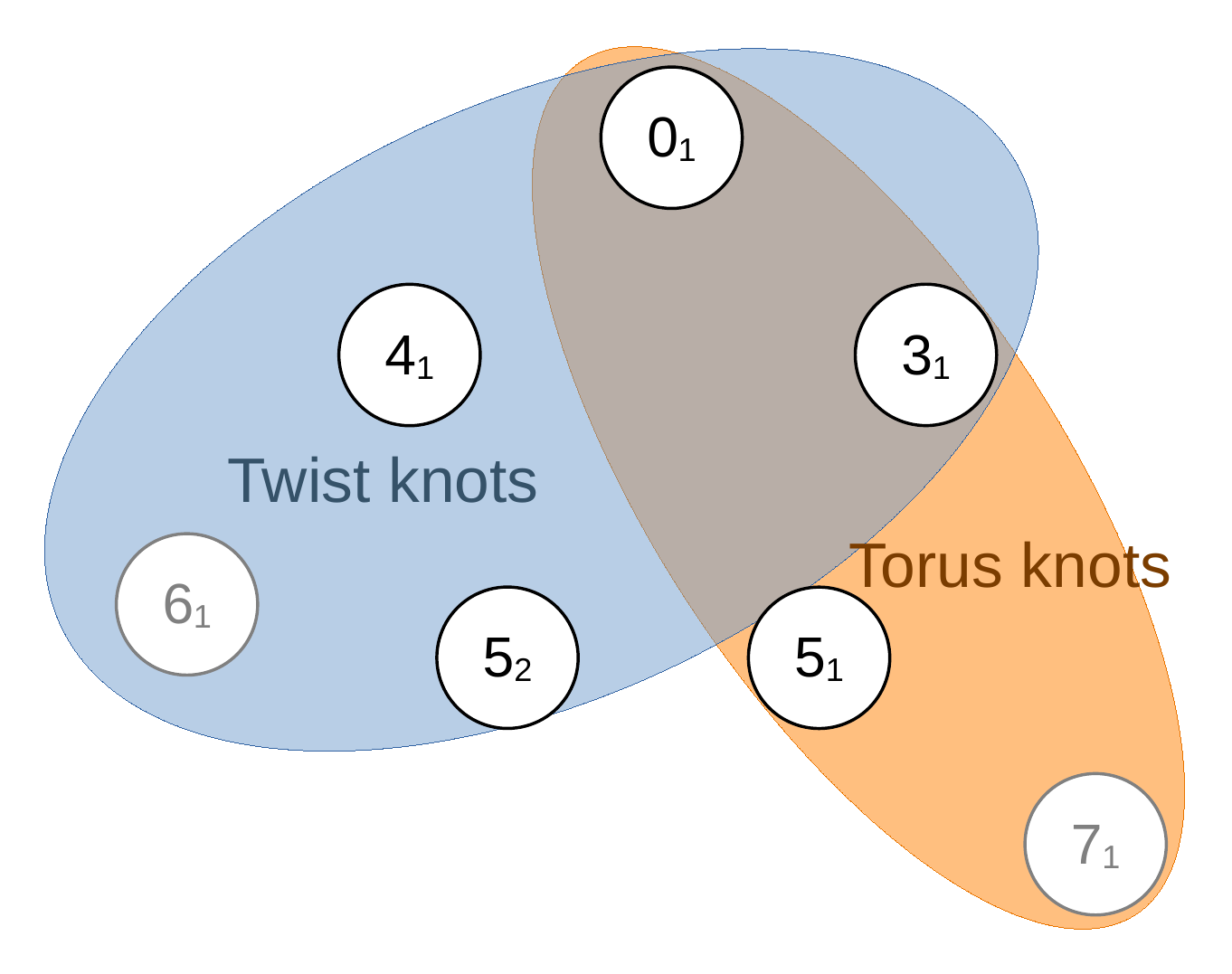}
\vspace{-1mm}
\caption{Sketch of families of knot types: simple torus knots and twist knots.
}
\label{fig:torus-twist}
\end{figure}

To further corroborate the indication that ML can, to a given extent, recognize topological properties that are common to a set of different knot types, we test our NNs on configurations having knot types never used for training. In particular, we test the LSTM on knotted rings with $k_c= 7_1$ (the torus knot with 7 minimal essential crossings) and $k_c=6_1$ (the twist knot next to $5_2$ in the complexity ladder). Some simple knot types and their families are sketched in Fig.~\ref{fig:torus-twist}.

Since the $6_1$ and $7_1$ knots were not included in the possible outputs of the NNs during training, here we are checking which, among the 5 knots learned by the NNs, are going to be associated with the novel knots.

On average we observe that for the best performing features such as coordinates, bonds and the set of 3 dihedrals, $95.6\%$ of rings with knot type $6_1$ is predicted to be a $4_1$ knot, i.e. the closest twist knot that has been used for training. Similarly, $98.7\%$ of the rings with knot type  $7_1$ is predicted as a $5_1$ knot. 

These results indicate that our NN is able to learn some generic topological properties of the knotted rings with a good  degree of accuracy. Notably, the ability to distinguish knots in their knot families works properly not only for knot types included in the training set but also for novel knots. In this respect, these validation tests suggest the fascinating possibility (to be further explored in the future) of having NNs not only as predictors of knot types included in the training set ($0_1,3_1,4_1,5_1,5_2$ in our case) but also as a general knot family indicator.

\section{Conclusions}\label{conclusions}

Our study confirms that machine learning techniques based on neural network architectures, when trained on proper sets of features, are able to identify knot types in confined rings, irrespective of their degree of compactness. 

Moreover, we have shown that a NN trained on rings of a given contour length $N$ can be used to recognize knots in rings of different lengths $N'$ if the corresponding features are first rescaled by a fine or coarse-graining procedure. 

Quite remarkably, even when the ML either fails to identify the right knot type or the analyzed configurations host unseen knots it nevertheless recognizes the correct topological family to which these knotted rings belong to. 

Finally, we have shown that the features that often provide the best performance are the ``raw"  information of the analyzed configuration such as its (rescaled) coordinates or bond directions. However, these features are not rotational invariant (they change when the ring is rotated in space), as opposed to the hosted knot type. In this respect, the use of such features does not follow the good practice of feeding NNs with inputs respecting the underlying symmetries of the system under study~\cite{noe2020machine}.
As a matter of fact, when computing energy functions of molecules, usually a set of distances and angles between atoms is considered. In our case, however, the chosen set of rotational invariant features such as distances, angles, and dihedral angles do not seem to provide the same level of accuracy obtained with the raw quantities, unless  features of larger dimensions are considered  (i.e. going from $3\times N$ to a $9\times N$ set of dihedral angles for each ring). Other rotational invariant features can be considered though. For instance in~\cite{sleiman2022geometric} it was suggested that the local density along the  ring (i.e. the average number of neighboring beads) could be a set of rotational and translational features that can faithfully capture the underlying topology of the rings. 

Clearly, a full understanding of how ML can recognize topological (i.e. non local) patterns within entangled closed curves and of which NN architecture or set of features can furnish the best performance
is still lacking. We hope our study will inspire and inform future investigations on this problem.

\acknowledgments{We thank Michele Caraglio and Davide Michieletto for useful discussions.
Funding from research grants `ORLA$\_$BIRD2020$\_$01' and 'BAIE$\_$BIRD2021$\_$01' of Universit\`{a} degli Studi di Padova is gratefully acknowledged.}

\appendix*\label{appendix}
\section{Convolutional Neural Networks}

\begin{figure*}[tbh!]
\centering
\includegraphics[width=.63\textwidth]{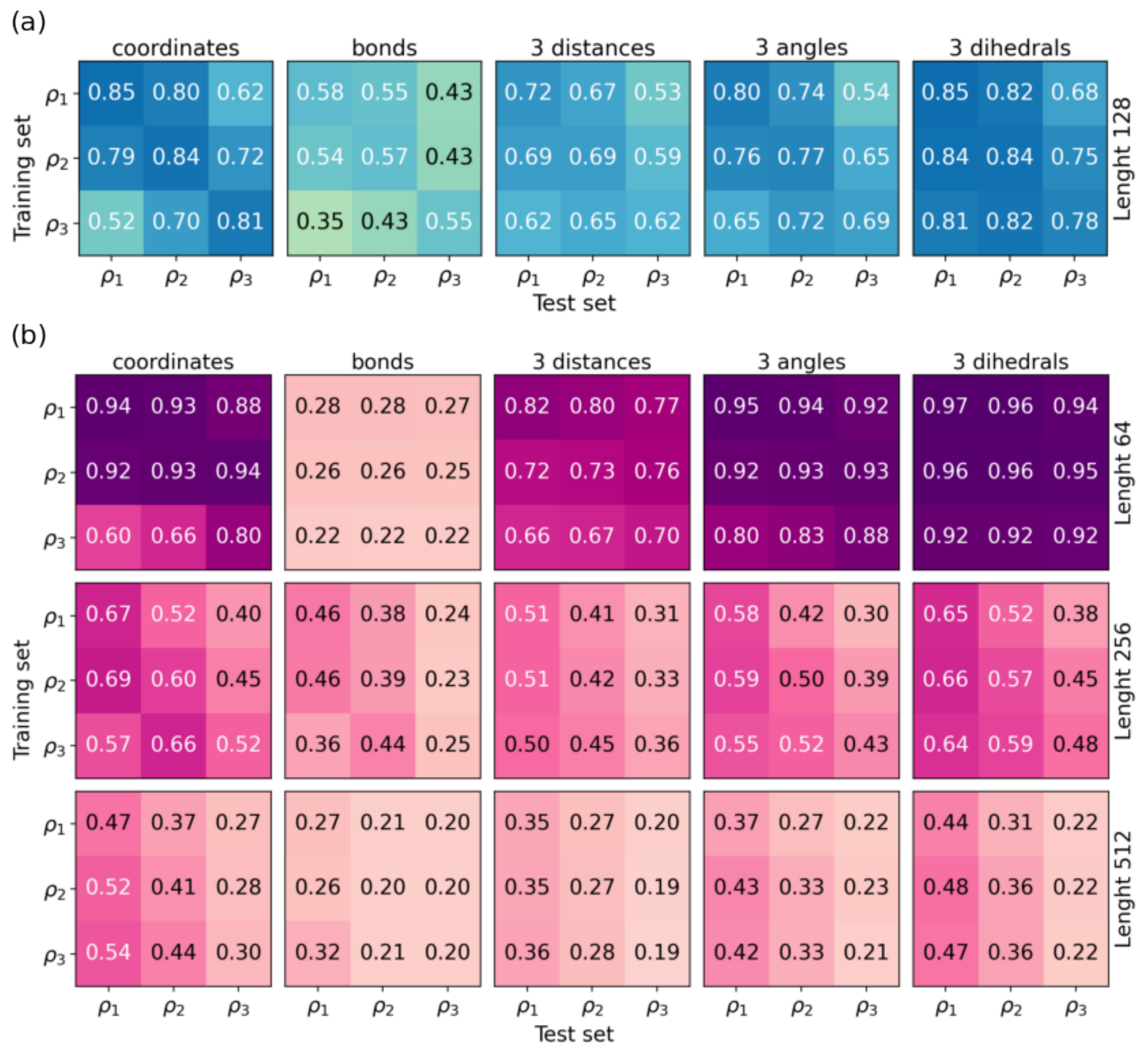}
\caption{For a CNN trained with various features (columns) of configurations with $N=128$:
(a) test accuracy for $N'=128$ for different training densities (left axis) and at different test densities (bottom axis);
(b) the same for $N'=64, 256$, and $512$ after rescaling of the configurations, as described in the text.
}
\label{fig:accCNN}
\vspace{-1mm}
\end{figure*}

CNNs are feed-forward neural networks whose architecture allows to automatically recognize patterns within images, sequences, etc., which are relevant for achieving a given task (classification, regression, object detection etc.). Usually, CNNs are made up of a series of convolutional blocks and a fully connected component, eventually returning the prediction. Each convolution block may include two main components: a convolution layer that processes the input signal through learnable filters and a pooling layer that reduces the dimension of the processed signal \cite{reviewML}. This architecture, in its 1-dimensional version, may be also applied to time series or sequences. Thus, we test its performance on polymer configurations (Fig.~\ref{fig:accCNN}) and we compare it with that of LSTM (Fig.~\ref{fig:acc}, Fig.~\ref{fig:acc2},  Fig.~\ref{fig:acc3}).

In general, we observe lower accuracy $A$ with respect to LSTMs, both for training length $N'=N=128$ and for other lengths $N'=64,256,512$ in their rescaled version. Moreover, we notice that the best performance refers to the feature set including 3 dihedrals while the worst one refers to bonds. This result suggests that for simpler ML models, such as the CNNs, more complex and processed features are needed to reach higher performance in knot recognition. 

The limit of the analysis via CNN may be due to the finite kernel size of filters in the convolutional layers. Correlated elements can be detected within filters but not outside them. Eventually, only a repetition of convolution on pooled (coarse-grained) layers might detect long-range correlations in a sequence of features, while this is naturally achieved with LSTMs.
To mild this problem, it seems useful to feed CNNs with complex functions of beads along the chain, such as dihedrals, angles, and distances. Nevertheless, LSTMs still produce higher accuracy, suggesting that the recurrent architecture is more appropriate for knot type prediction in polymer configurations, even with simpler features, such as coordinates and bonds.


%

\end{document}